\documentclass[aps,twocolumn,superscriptaddress,floatfix,prl]{revtex4-2}
\usepackage{float}
\usepackage{mathrsfs}
\usepackage{graphicx}
\usepackage{amsmath,amssymb}
\usepackage{multirow}
\usepackage{subfigure}
\usepackage{braket}
\usepackage{bm}
\usepackage{color}

\begin{document}
\title{Soliton States From Quadratic Electron-Phonon Interaction }

\author{Zhongjin Zhang}
\affiliation{Department of Physics, University of Massachusetts, Amherst, Massachusetts 01003, USA}
\author{Anatoly Kuklov}
\affiliation{Department of Physics \& Astronomy, CSI, and the Graduate Center of CUNY, New York 10314, USA}
\author{Nikolay Prokof’ev}
\affiliation{Department of Physics, University of Massachusetts, Amherst, Massachusetts 01003, USA}
\author{Boris Svistunov}
\affiliation{Department of Physics, University of Massachusetts, Amherst, Massachusetts 01003, USA}
\affiliation{Wilczek Quantum Center, School of Physics and Astronomy and T. D. Lee Institute, Shanghai Jiao Tong University, Shanghai 200240, China}

\bibliographystyle{apsrev4-2}

\begin{abstract}
We present the first numerically exact study of self-trapped, a.k.a.  soliton, states
of electrons that form in materials with strong quadratic coupling to the phonon 
coordinates. Previous studies failed to observe predictions 
based on the variational approach in continuum space because soliton states
form only when system parameters are taken to the extreme limit. 
At the variational level, we establish that finite-radius solitons emerge 
through the weak first-order transition as the coupling strength is increased, 
and subsequently collapse to the single-site state through strong first-order 
transition. Both transitions transform into smooth crossovers between the light 
and heavy polaron states in the full quantum treatment. The most surprising 
effect not observed in any other polaron model is non-monotonic dependence 
of the soliton effective mass and the residue at strong coupling. 
\end{abstract}
\maketitle

\textit{Introduction}.
For decades, studies of polarons---electrons renormalized by their coupling to
lattice vibrations \cite{Landau1,Landau2,Landau3}---were focused on the linear density-displacement electron-phonon interaction (EPI): 
\begin{equation}
H_{\rm int} = \sum_{{\mathbf k} {\mathbf q} \sigma \alpha } V_{{\mathbf q} \alpha}
c_{{\mathbf k}- {\mathbf q} \sigma}^{\dagger} c_{{\mathbf k} \sigma}^{\;}  \left[ b_{{\mathbf q} \alpha}^{\dagger} + b_{-{\mathbf q}\alpha}^{\;} \right] ,
\label{linear}
\end{equation}
where $\alpha$ is the phonon branch index
(we use standard notations for creation and annihilation operators for electrons $c_{{\mathbf k}, \sigma}^{\dagger}, c_{{\mathbf k} \sigma}^{\;} $ and phonons $b_{{\mathbf q} \alpha}^{\dagger} , b_{-{\mathbf q}\alpha}^{\;}$ in momentum representation). 
The most popular models were
the local Holstein \cite{Holstein1,Holstein2} and the non-local Frohlich \cite{Frohlich1,Frohlich2} models with $V_{{\mathbf q} \alpha} = {\rm const}$ and $V_{{\mathbf q} \alpha} \propto 1/q$,
respectively (for the recent review see Ref.~\cite{Franchini}).
More recently, researchers started exploring alternative interaction mechanisms based on the electron hopping amplitude dependence on atomic displacements \cite{Mona1, Mona2,Scalettar,Chao1,Chao2,Sous1,Chao3} that bring   
an additional dependence of the vertex function in (\ref{linear}) 
on the incoming electron momentum, $V_{{\mathbf q} \alpha} \to V_{{\mathbf k},{\mathbf q} \alpha}$, but remain linear in the phonon coordinates. 
The interest in new models was motivated by their unusual properties and the possibility of having light but compact bi-polarons
(bound states of two electrons) with high superconducting transition temperature 
\cite{Chao2,BoseTc1}.

Much less explored and understood remained properties of polarons
with quadratic EPI: 
\begin{equation}
H_{\rm int} = g_2 \frac{\Omega}{4} \sum_{\mathbf i} n_{{\mathbf i}} [ b_{\mathbf i}^{\dagger} + b_{\mathbf i}^{\,}]^2 .
\label{quadratic}
\end{equation}
Here $\Omega $ is the local oscillator frequency in the absence of coupling; it changes to $\tilde{\Omega } = \sqrt{1+g_2n_i}\, \Omega$ 
when the site is occupied. Note crucial dependence of the model's properties on the sign of $g_2$ and, in particular, the instability taking place at $g_2< -1$.

An intriguing regime emerges when the two limits, $\Omega \to 0$ and $g_2 \to +\infty$, conspire to preserve finite $\tilde{\Omega}$. 
Indications of the importance of this regime were found in several materials: 
doped manganites \cite{X2materials1}, halide perovskites \cite{X2materials2},
and quantum paraelectrics \cite{X2materials3,X2materials4}.
The soft vibration modes in these materials are
transverse optical phonons for which the linear EPI is suppressed
in the long-wave limit \cite{soft1,Kuklov,Kuklov2,Gogolin,Gogolin2};
similar physics takes place in optically pumped systems \cite{Andy1,Andy2}.
Early suggestions that bi-phonon exchange could be an important pairing
mechanism at low doping \cite{soft1} were recently revisited and used
to explain superconductivity in SrTiO$_3$ \cite{soft2,soft3,soft4}.
However, dealing with non-linear couplings theoretically beyond perturbation theory has been challenging. The original work \cite{Kuklov,Kuklov2,Gogolin,Gogolin2} was based on the
variational solution for large-radius soliton states in continuum.
The momentum average approximation \cite{Mona0} was used to
study a combination of linear and nonlinear EPI in Refs.~\cite{Mona00,Mona000}.
More recently, nonlinear EPI effects were investigated in 
Refs.~\cite{detMC1,detMC2,detMC3} using the determinant
Monte Carlo method \cite{detMC4} for finite 2D systems  
at high electron density and finite temperature.
Finally, the interplay between the linear and quadratic EPI
in continuum was studied in Ref.~\cite{Tempere} by the variational Feynman’s path-integral method \cite{Feynman}.

The situation has changed with the development of the
numerically exact $x$-representation Monte Carlo (XMC) technique for polaron problems with arbitrary non-linear density-displacement and
arbitrary sign-preserving hopping-displacement interactions \cite{Xrep}. It was used to obtain 
the first precise results for quadratic EPI (\ref{quadratic}) at strong coupling \cite{X2polaron}, but the study failed to observe soliton states predicted by Ref.~\cite{Kuklov,Kuklov2} (in what follows we will call them ``kuklons'') despite considering relatively
small adiabatic parameter $\gamma = \Omega/W = 1/48$, where $W=12 t$ is the bandwidth of the 3D tight-binding model on the cubic lattice with the nearest-neighbor hopping amplitude $t$. Thus, no progress on the soliton problem was made for more than three decades, and it remains unknown how and under what conditions these states form and what their basic properties are.  

In this Letter, we first expand the variational analysis of the adiabatic limit to reveal how kuklons form and then collapse 
in first-order phase transitions. Next, we present results of the numerically exact studies of kuklons by the XMC method in the parameter  regime that is orders of magnitude beyond the limitations of all 
other known schemes: $\gamma = 1/600$ and $\gamma = 1/2400$ 
with $g_2$ as large as $10^8$. In the full quantum solution, the first-order transitions are transformed into smooth crossovers with the most unusual (not observed in any other polaron problem)  
non-monotonic dependence of the effective mass and $Z$-factor 
on $g_2$ at strong coupling. 

\textit{Hamiltonian, effective adiabatic model, and methods}.
The full system's Hamiltonian on the simple cubic lattice reads 
\begin{equation}
            H =6t-t\sum_{\braket{ij}}c_i^\dagger c_j+\Omega\sum_ib_i^\dagger b_i + H_{\rm int} \, .
\label{Hamiltonian}
\end{equation}
[The phonon energy is counted from the ground state of the unperturbed harmonic oscillator.] In the adiabatic, $\gamma \to 0$, limit one
takes advantage of the fact that electrons are much faster than phonons
and the contribution of the latter to energy depends solely on the average electron density distribution \cite{Landau1,Landau2,Kuklov,Kuklov2}.
For model (\ref{Hamiltonian}), these standard considerations lead to the following energy functional to be minimized: 
\begin{equation}
          E =6t -t \sum_{\braket{ij}}\psi_i^*\psi_j+\frac{\Omega}{2}\sum_i\left(\sqrt{1+g_2\left|\psi_i\right|^2}-1\right) .
\label{H_dis} 
\end{equation} 
Here $\psi$ is the normalized electron wave function (real for the ground state), 
and the last term 
is the sum of ground-state energies for each oscillator. 
Given that oscillator frequencies increase with $g_2$, the adiabatic condition for the polaron of radius $R$ is satisfied if $\sqrt{R^{-3}g_2}\Omega \ll t/R^2$ or $g_2 \ll (t/\Omega)^2/R$ 
(the lattice constant $a=1$ serves as the unit of length). 
For large-radius kuklon, we also consider a continuous counterpart of 
Eq.~(\ref{H_dis}) with $m=1/2t$:
\begin{equation}
         E =\int {\rm d}^3r \left[\frac{1}{2m}\left|\nabla\psi\right|^2+\frac{\Omega}{2}\left(\sqrt{1+g_2\left|\psi\right|^2}-1\right)\right] .
\label{H_cont}
\end{equation} 

Minimization of Eq.~(\ref{H_dis}) is achieved by the
gradient decent method when at each stage 
the wave function is first transformed according to 
$ \tilde{\psi} =  \psi -\epsilon \nabla_\psi H$ and then normalized to unity.
This step is rejected and the value of $\epsilon$ is decreased by a factor of two
if $E[\tilde{\psi}]>E[\psi ]$; otherwise it is accepted. 
For optimal solutions we record their energies and root mean square
radii, $R = \sqrt{\braket{r^2}}$, where  
$\braket{r^2} =\sum_i r_i^2 \psi_i^2$. We ensure that all 
finite-size effects are exponentially small for the polaron solutions
presented in this paper (realistically, one can work with about 
$300^3$ sites after utilizing system symmetries).
        
Numerically exact solutions of (\ref{Hamiltonian}) are obtained 
using recently developed XMC method based on the lattice path integral for the electron and coordinate representation for harmonic oscillators
(see Refs.~\cite{Xrep,X2polaron} for complete description). Key polaron properties such as dispersion relation, $E_p$, 
ground-state energy, $E=E_{p=0}$, effective mass, $1/m^*=d^2E_{p=0}/dp^2$, and the quasiparticle residue, $Z=Z_{p=0}$, are obtained from simulations of the polaron Green's function, $G_p(\tau)$, and its 
asymptotic behavior, $G_p( \tau)  \to Z_p {\rm e}^{-E_p \tau}$, 
in the $\tau \to \infty $ limit.

\textit{Variational analysis}.  
The formation of kuklons is driven by competition between 
the hopping term favoring delocalized states and EPI repulsion 
preferring localized states due to sub-linear dependence of the interaction energy on density at strong coupling [last terms in (\ref{H_dis}) and (\ref{H_cont})]. Since delocalized states 
correspond to $\psi_i \to 0$, their variational energy readily 
follows from Eq.~(\ref{H_dis}) in this limit: $E_0=g_2\Omega /4$. 
 
Localized continuous solutions at strong coupling were obtained 
in Ref.~\cite{Kuklov,Kuklov2}. If the last term in (\ref{H_cont}) 
is approximated as $\Omega\sqrt{g_2}|\psi|/2 $, then energy 
minimization reduces to the solution of the radial eigenvalue 
equation, 
$-d^2 \psi /dr^2 + m\Omega\sqrt{g_2}/2 - \lambda \psi = 0 $, with the solution 
\begin{equation}
    \psi(r)=\sqrt{\frac{3}{10\pi R_A^3}}\left[1-\frac{R_A\sin\left(x_0r/R_A\right)}{r\sin\left(x_0\right)}\right],
\label{kuklon}
\end{equation}
for $r<R_A$ and $\psi(r)=0$ for $r\ge R_A$, 
where $x_0\approx 4.49$ is the lowest positive root of 
$\tan x = x$ and $R_A$ is the cut-off radius,
$R_A = [6x_0^4/(5\pi g_2\Omega^2m^2)]^{1/7}$. 
The root mean-square radius of this solution is given by 
$R=(11/25-49/10x_0^2)^{1/2} R_A \approx 0.444 R_A$ and  
its energy
\begin{equation}
   E_A = \frac{7}{10}\left(\frac{5\pi x_0^3}{6}\right)^{2/7}\frac{\left(g_2\Omega^2\right)^{2/7}}{m^{3/7}} 
   \approx 3.34 \frac{\left(g_2\Omega^2\right)^{2/7}}{m^{3/7}}.
\label{EA}
\end{equation}
falls below $g_2\Omega/4$ for $g_2 > 37.7 (m\Omega )^{-3/5} $.

Thus, on the one hand, kuklons form at strong coupling $g_2 \gg 1$. 
On the other hand, this coupling needs to be small enough to 
ensure adiabaticity of the state and its large radius because 
at $g_2 > (t/\Omega)^2$ both conditions are violated.
At $g_2 \gg (t/\Omega)^2$ the polaron state is well described 
by the so-called atomic limit (AL) when the electron changes 
frequency only of the harmonic oscillator at the occupied site, see Ref.~\cite{X2polaron}. Physically, this is a completely different 
state characterized by $R\approx 0$ and $E_{\rm atom} \approx \frac{\Omega}{2}(\sqrt{1+g_2}-1)$. 
\begin{figure}[ht]
			\begin{center}
			\includegraphics[scale=0.35]{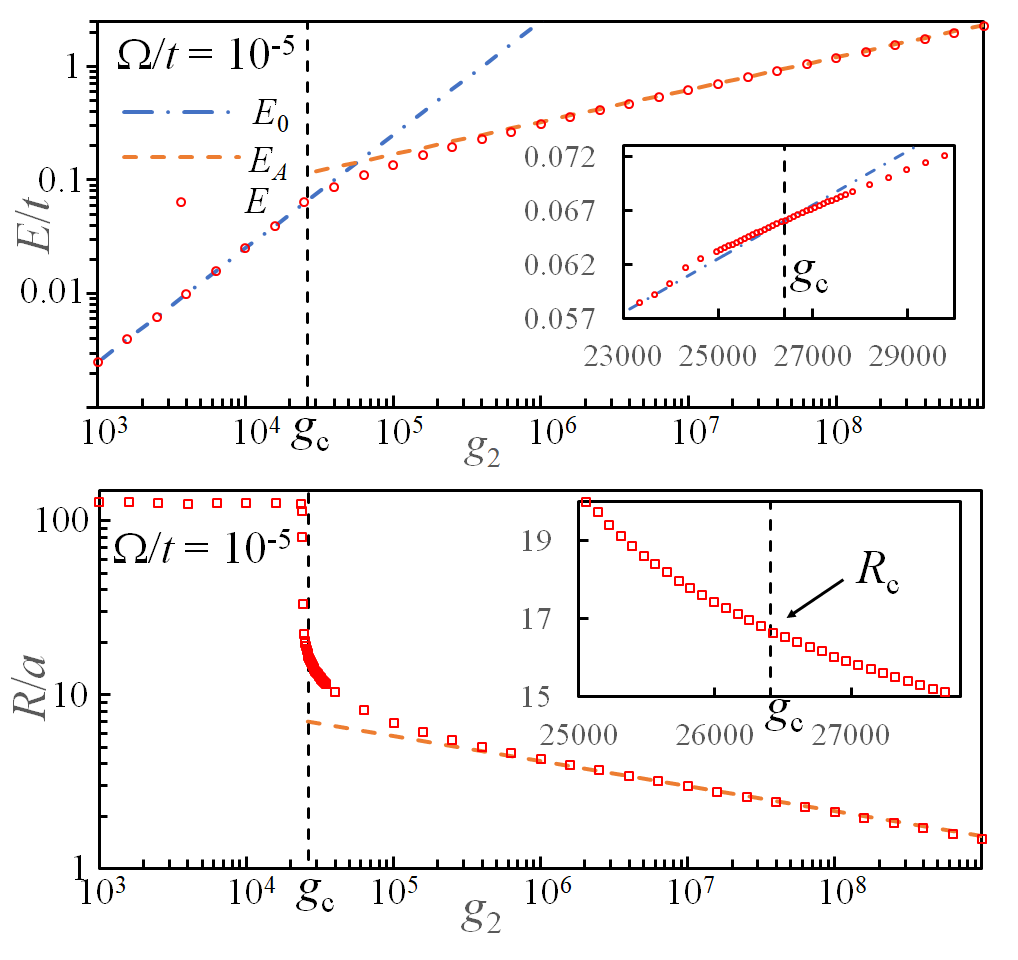}
                \end{center}
\caption{ 
(color online) Energy (upper panel, red circles) and radius 
(lower panel, red squares) of lattice variational states as functions 
of $g_2$ for $\Omega/t = 10^{-5}$. 
Dashed (orange) lines are the asymptotic continuous solutions, 
see Eq.~(\ref{EA}) and the text above. 
Dash-dotted (blue) line is energy of the delocalized state. 
Insets: Energy and radius vs. $g_2$ near the critical point.  }
\label{Evsg201}
\end{figure}
 
In a more precise treatment, the last term in (\ref{H_cont}) 
is dealt with ``as is" and the cutoff at $R_A$ gets replaced 
with gradual exponential decay of the wavefunction with distance; this feature is important for correct description
of the transition between the localized and delocalized states. The corresponding numerical analysis is presented below. In the absence of
hard cutoff, we have to resort to the root mean square radii
for defining the polaron size. 

To study properties of kuklons away from the asymptotic 
adiabatic limit $(\gamma \to 0,\, R \to \infty )$---how they first 
form and then transition to the AL state, we resort to the exact
minimization of the lattice functional (\ref{H_dis}) by gradient 
decent. Using $t = 1$ as the energy unit, we first consider the case
of extremely small phonon frequency $\Omega/t= 10^{-5}$, see Fig.~\ref{Evsg201}. 
For $g_2\gg g_c$ ($g_{c} = 26406$), both $E$ and $R$ 
are accurately described by the asymptotic continuous solutions. 
However, this is no longer the case on approach to $g_c$, because exact 
variational solutions remain stable at significantly smaller 
(by a factor of two) values of $g_2$ as their volume undergoes a  
rapid expansion (by nearly an order of magnitude). This is a clear indication 
that the hard-cutoff approximation is not reliable near the critical point.

Close examination of the $|g_2-g_c|/g_c \ll 1$ region, see insets in Fig.~\ref{Evsg201}, reveals that within the variational treatment, kuklons with finite radius $R_c$ 
emerge through a weak first-order transition. Indeed, for $g_2$ slightly smaller than $g_{c} $, we detect metastable localized solutions (obtained by starting from an initial wave function with a small radius), which subsequently disappear at $g_2 < 0.92 g_{c}$. 
The phase transition point can be located very accurately from the intersection of the $E$ and $E_0$ curves. 
\begin{figure}[ht]
			\begin{center}
			\includegraphics[scale=0.35]{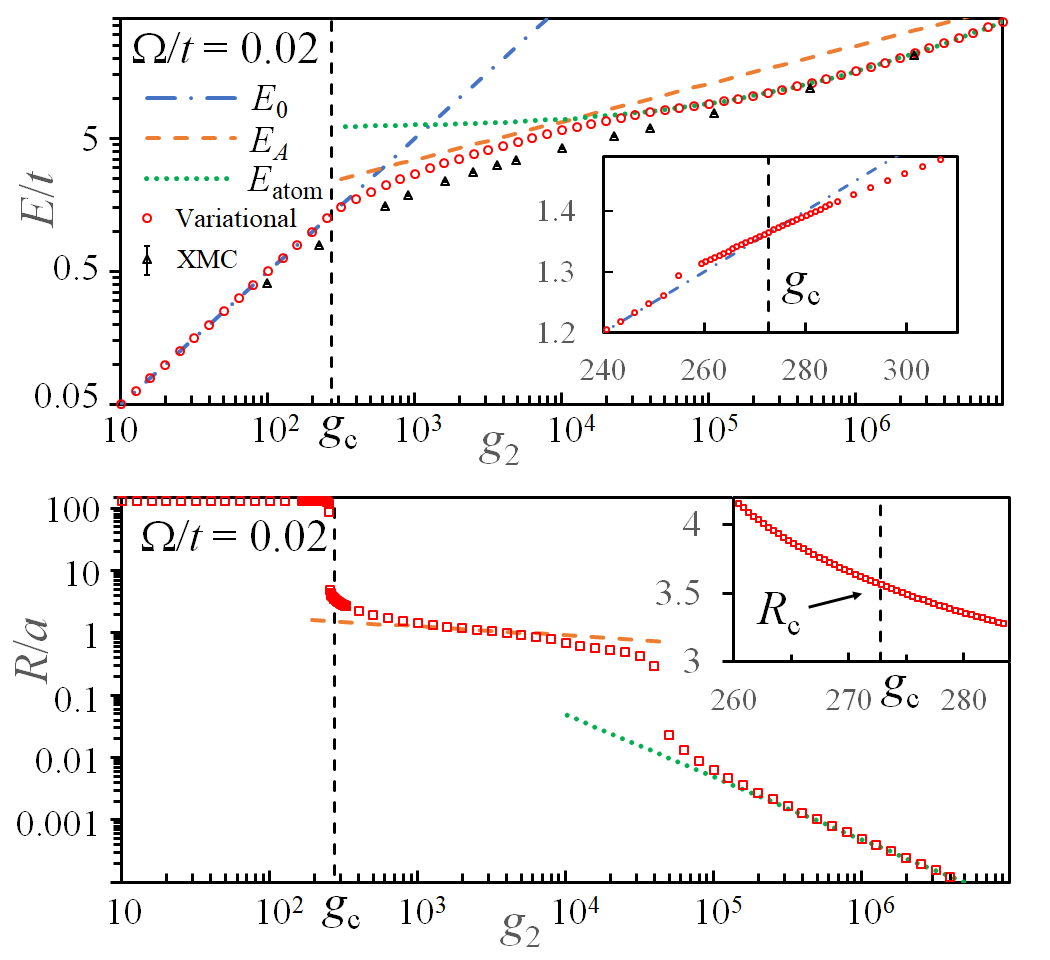}
                \end{center}
\caption{(color online) Upper panel: Energies of variational states (red cicles)
in comparison with XMC results (black triangles) for $\Omega = 0.02$. Lower panel: variational state radii (red squares). 
Dashed (orange) lines are predictions of the asymptotic continuous solutions. 
Dash-dotted (blue) line is the energy of the delocalized variational state $E_0$. 
Dotted (green) line is the atomic limit energy $E_{\rm atom}$. 
Inset: Variational $E$ vs. $g_2$ near the critical point. }
\label{Evsg202}
\end{figure}

\begin{figure}[ht]
\begin{center}
	\includegraphics[scale=0.24]{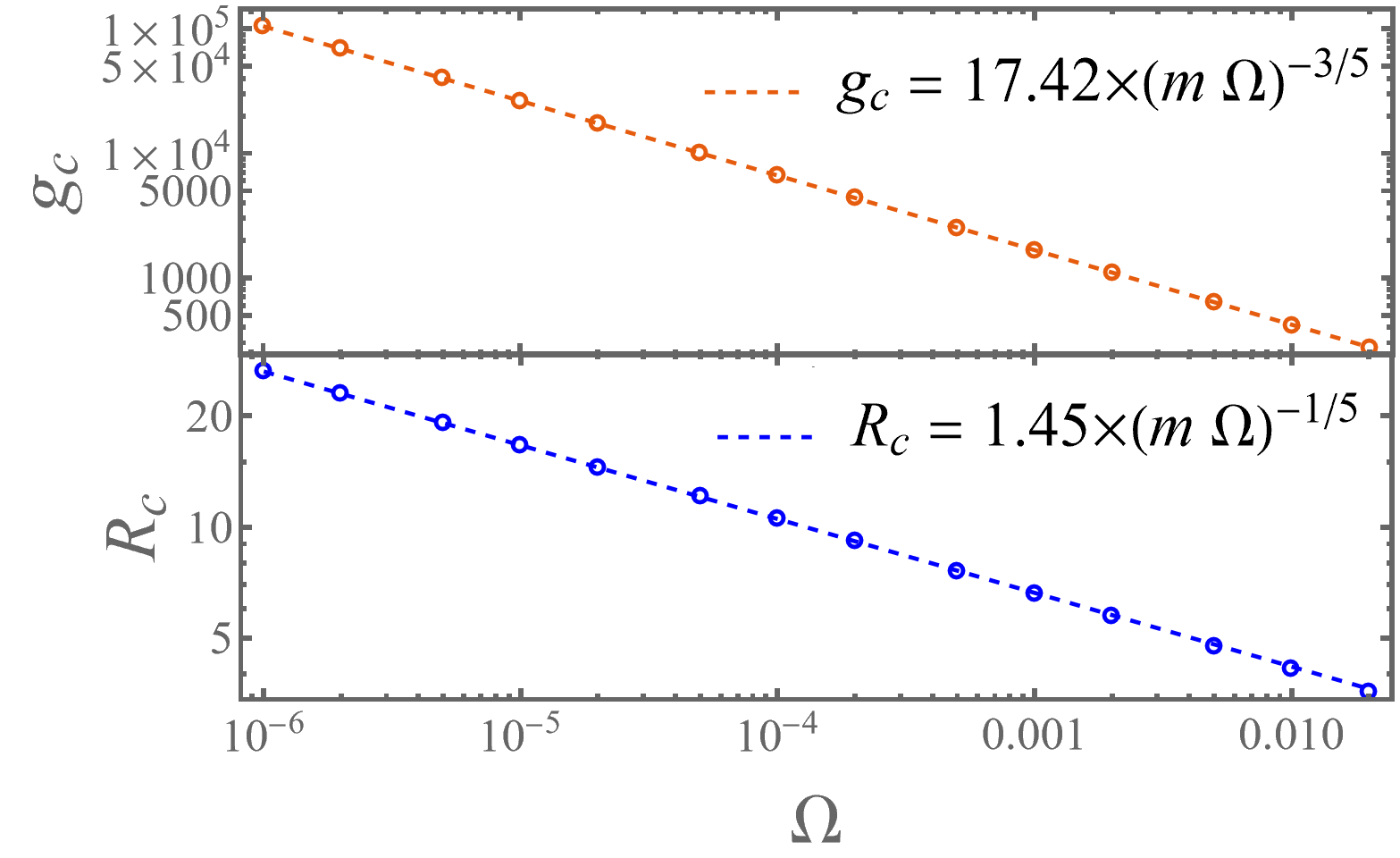}
\end{center}
\caption{Critical coupling and radius of kuklons in continuum
as a function of $\Omega$. Both show clear power-law dependence 
(dashed lines) with expected exponents.}
			\label{g2vsOmega}
\end{figure}

In continuum, the energy functional is invariant under the scaling transformation, $r\rightarrow  b r$, $g_2\rightarrow b^{3} g_2$,
$\Omega \rightarrow b^{-3} \Omega$, and  $m\rightarrow b^{-2} m $.
Moreover, if $m$ and $\Omega$ are changed while keeping the adiabatic parameter $m\Omega$ constant, the functional is simply 
multiplied by a factor $\tilde{\Omega}/\Omega $. Thus, upon proper scaling, we  are left with only one free parameter: $g_2$.  
These considerations imply that the critical coupling and radius 
scale as $g_{c}\propto (m \Omega)^{-3/5}$ and $R_c \propto (m \Omega)^{-1/5}$. On a lattice, this scaling is expected to fail
when $R_c \sim a$. Data in Fig.~\ref{g2vsOmega} demonstrate the accuracy 
of this prediction for large critical sizes.  
 
\begin{figure}[ht]
			\begin{center}
                \subfigure {\includegraphics[scale=0.2]{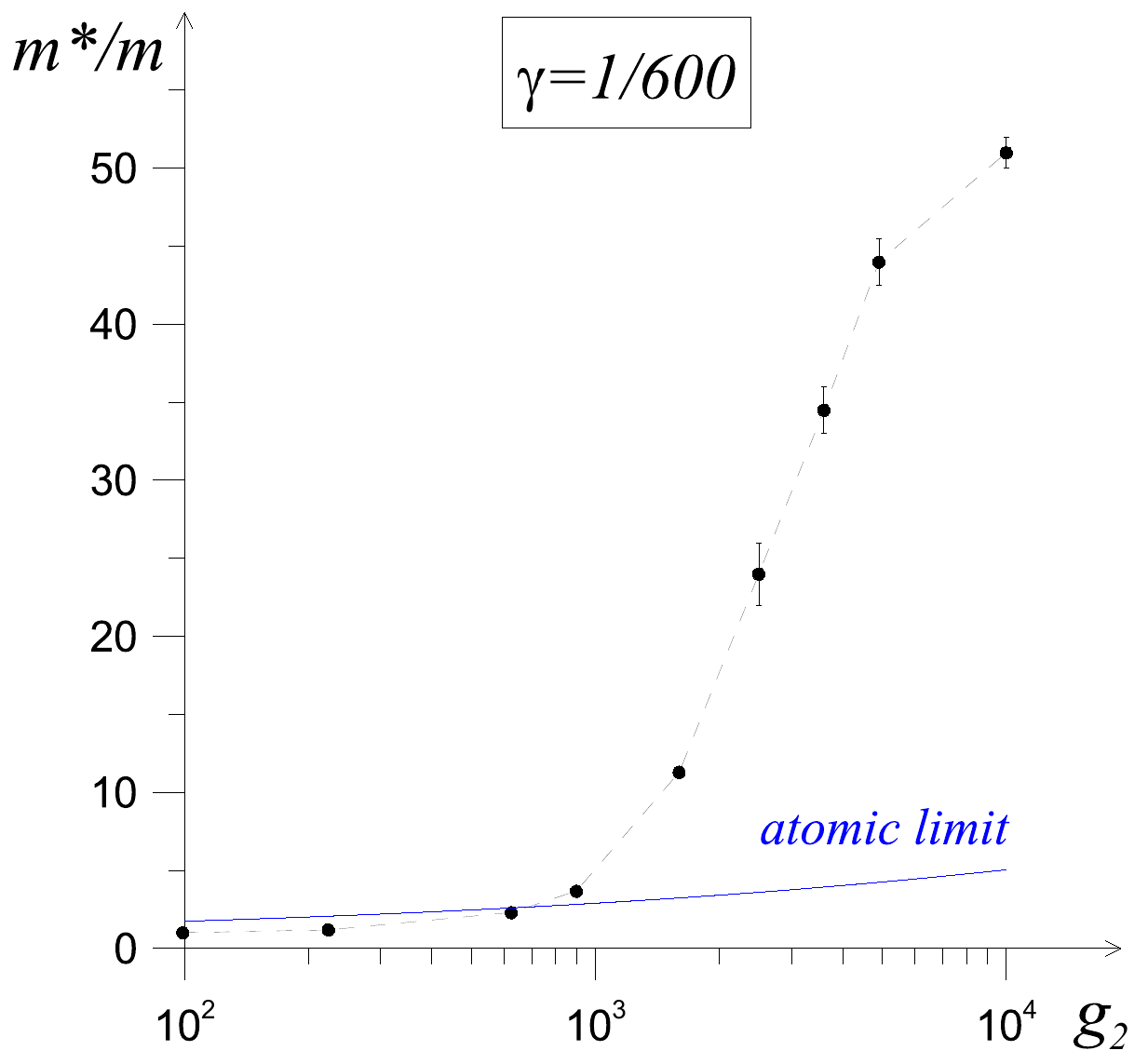}}
                \subfigure {\includegraphics[scale=0.2]{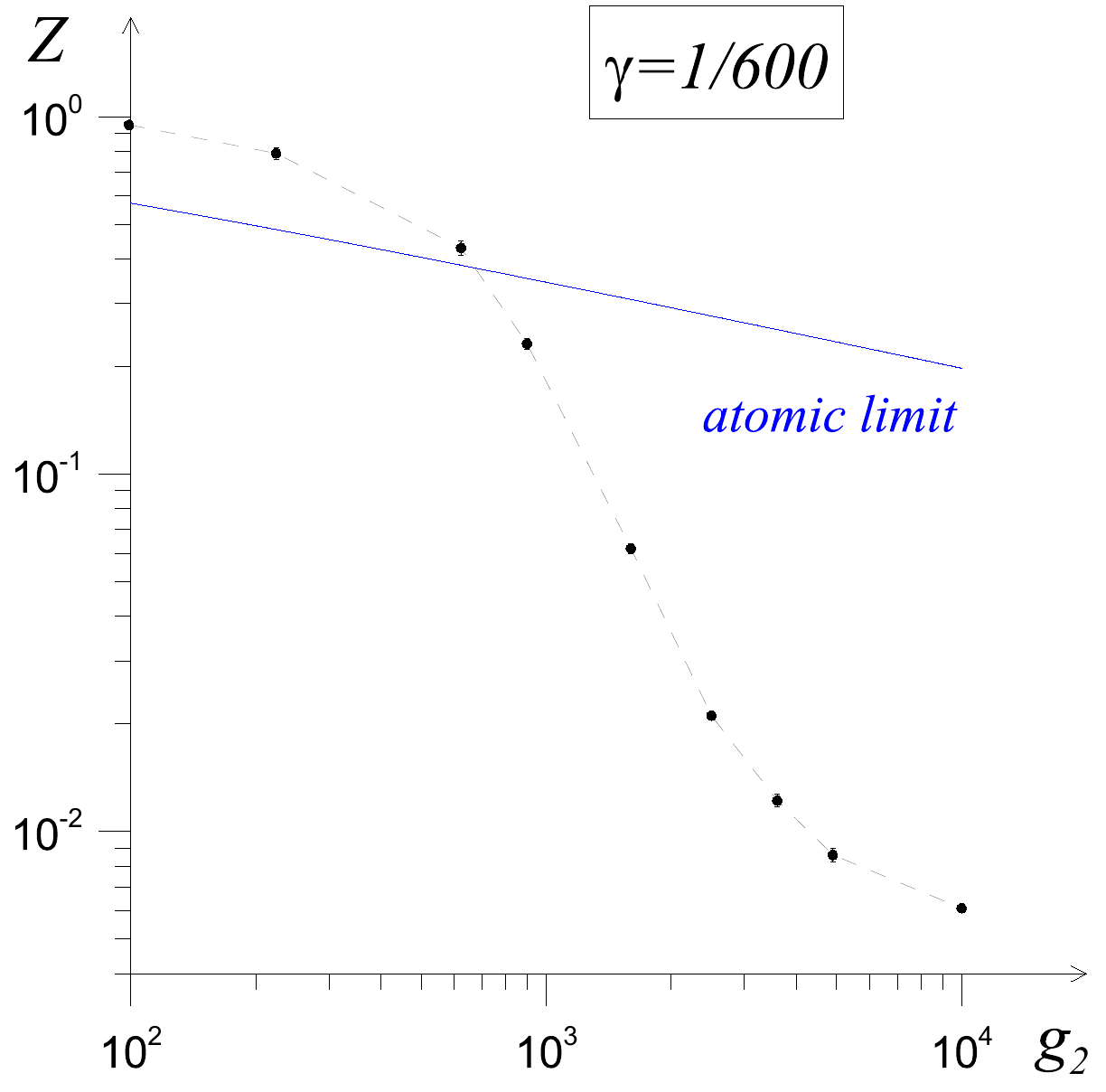}}
                \end{center}
\caption{(color online) Effective mass and $Z$-factor for $\Omega/t=0.02$
as functions of $g_2$ from the XMC method. Solid lines are results expected in the atomic limit}
\label{Effectivem}
\end{figure}

\textit{Quantum solution.} For a less extreme but still small adiabatic ratio $\gamma = 1/600$ 
we are posed to compare variational and full quantum solutions, see Fig.~\ref{Evsg202}. 
As expected, the first-order transition at $g_c=272.7$ predicted by (\ref{H_dis})
is replaced with smooth quantum crossover. 
Despite very small value of $\gamma$ the exact energy is still 
significantly different from its variational counterpart across the crossover region. 
This surprising outcome finds its explanation in relatively small kuklon sizes 
predicted by (\ref{H_dis}): 
about four lattice spacings at $g_c$ and quickly shrinking to $R \sim a$ in a broad 
parameter range. Quantum effects are expected to be pronounced at the lattice scale
not to mention that for $R\sim a$ the adiabatic condition also becomes questionable.
Nevertheless, it is clear that at $g_2 \sim 10^3$ the polaron state undergoes a radical transformation from perturbative plane-wave state with slightly renormalized effective 
mass and $Z\approx 1$ to a state with $m^*/m$ much larger and $Z$-factor much smaller
than what is expected in the AL, see upper panels in Fig.~\ref{Effectivem}. 
These results imply that the electron is ``dressed'' by oscillator excitations  
distributed over multiple sites, i.e. it forms a kuklon.
 
We observe that parameters required for formation of large-radius 
kuklons are extreme and well beyond the limits of applicability of any other 
unbiased method. For XMC, simulations of $\gamma < 0.001$ are also very 
challenging in the regime of large effective masses, $m^*/m \gtrsim 1000$, and exponentially small $Z$-factors. This explains why previous numerical studies failed to see them. 

\begin{figure}[ht]
			\begin{center}
                \subfigure {\includegraphics[scale=0.2]{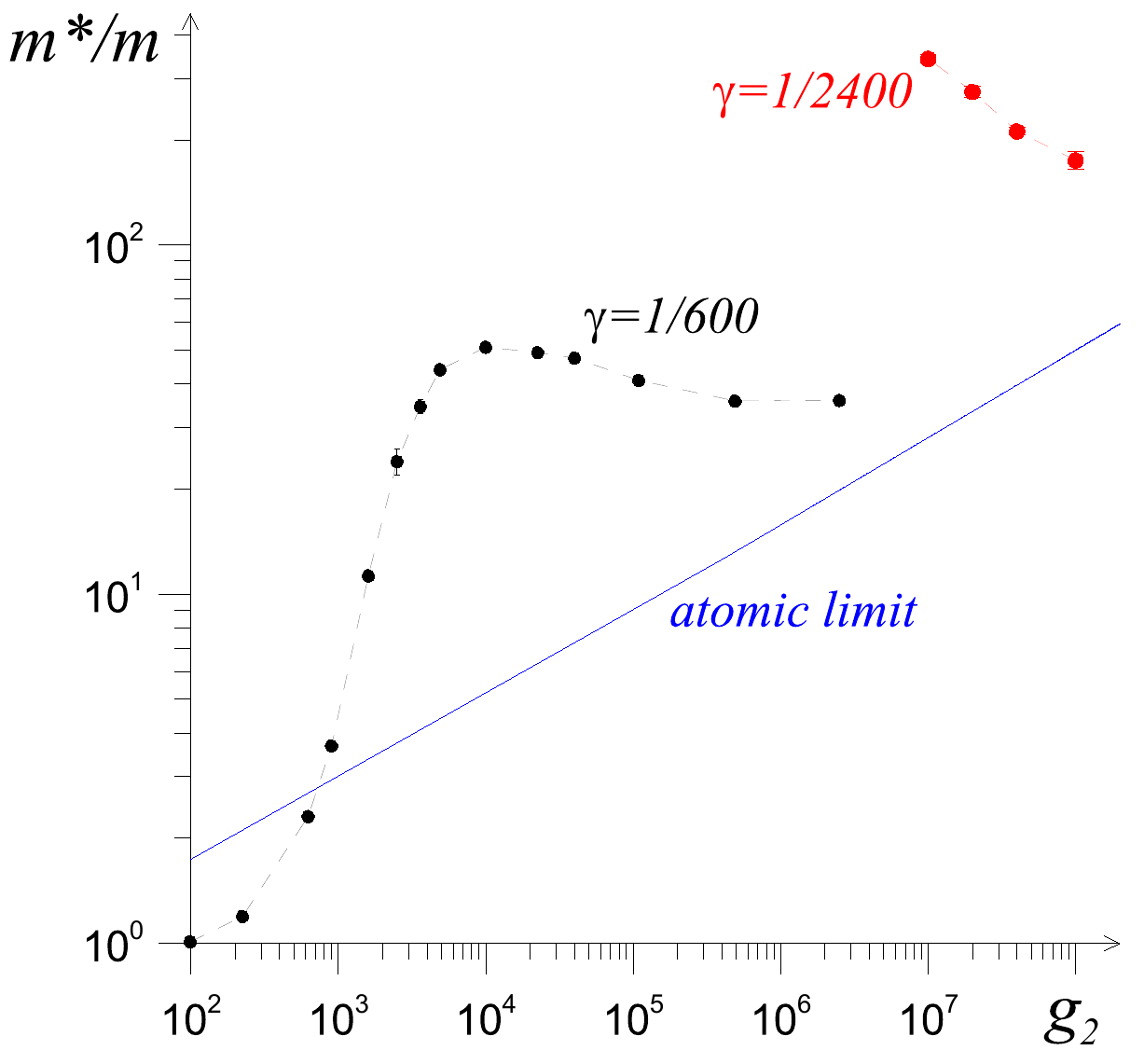}}
                \subfigure {\includegraphics[scale=0.2]{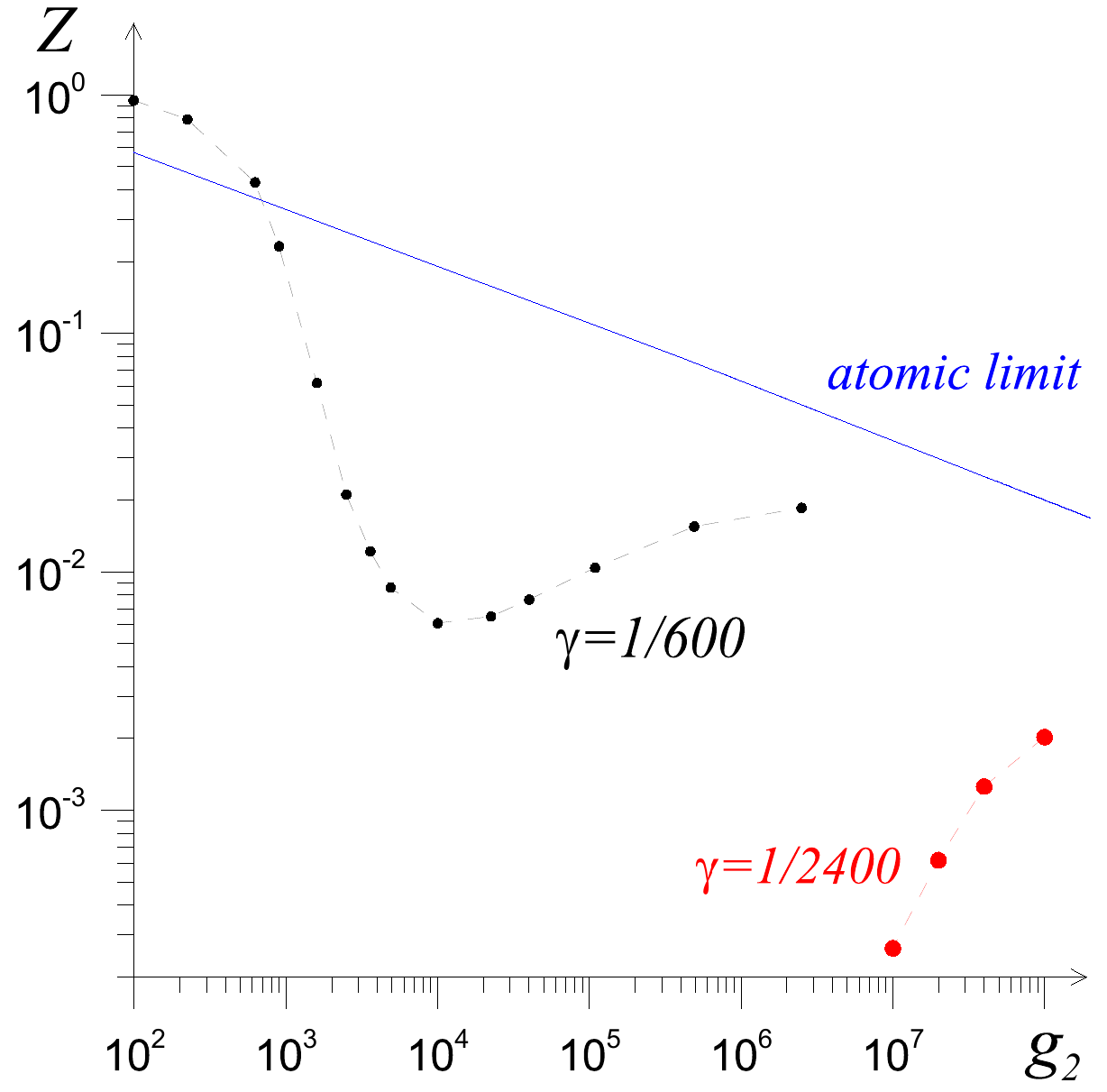}}
                \end{center}
\caption{(color online) Effective masses and $Z$-factors for $\Omega/t=0.02$ and $\Omega/t=0.005$ as functions of $g_2$ over a much broader range that in
Fig.~\ref{Effectivem}. Solid lines are results expected in the atomic limit.}
\label{Effectivem2}
\end{figure}
 
\textit{Collapse to the atomic limit. Lighter polarons at stronger coupling.}
Once kuklons form, their radius gradually decreases with coupling and at  
$g_2 \sim (t/\Omega)^2$ both lattice effects and non-adiabatic corrections 
come into play. Naive expectations are that AL is nothing but the 
end of the smooth monotonic evolution of the kuklon state. This turns out to be 
incorrect and in a rather dramatic fashion. Figure \ref{Evsg202} shows that 
the lattice functional features a second transition at large $g_2$, this time 
between the kuklon and AL states. It can be understood analytically 
by restricting analysis to just two wave function components near 
the transition point: $\psi_0 \simeq 1$ at the center and $\psi_1 \ll 1 $ at the six nearest neighbour sites (other components are proportional to higher powers of $\psi_1$).
By assuming that $\psi_1^2 g_2 \gg 1$ and using the normalization condition, $\psi_0^2 =1-6\psi_1^2$, we arrive at the energy dependence on $\psi_1$
\begin{equation}
    E/t = 6\left[6\psi_1^3-\chi\psi_1^2+2(\chi-1)\psi_1\right]+ {\rm const} 
\end{equation}
that features a first-order transition when $\chi=\sqrt{g_2}\Omega/(4t)=\chi_c$, with 
$\chi_c\approx 1.022$. 
At this level of description, the transition is from small but non-zero $\psi_1$ at 
$\chi < \chi_c$ to $\psi_1=0$ at $\chi > \chi_c$. In terms of the coupling constant, the transition takes place at $g_2 > (4t/\Omega)^2 = 9/\gamma^2$. 
[The above consideration is valid only in the $\gamma \to 0$ limit; more precise treatment 
with terms $\propto \chi/(g_2\psi_1)$ included shows that the discontinuous 
transition to the AL takes place only for $\gamma < 0.0015$.]  

While there is no reason to trust variational results for compact states, 
they do point to the possibility of having a much faster crossover to the AL
than what is expected in terms of slow power-low evolution $R \propto g_2^{-1/7}$.
Given an order of magnitude difference between the effective masses and $Z$-factors of
the kuklon and AL states, see Fig.~\ref{Effectivem}, monotonic power-law 
transformation of one state into another would require enormous values of $g_2$. 
But if this transformation takes the form of a more rapid crossover 
then $m^*$ and $Z$ must exhibit non-monotonic dependence on $g_2$ at strong 
coupling. Such highly counter-intuitive behavior was never reported for any 
polaron problem. By extending XMC simuations to much larger values of $g_2$ for 
$\gamma=1/600$ and performing additional simulations for $\gamma=1/2400$ 
at strong coupling we do observe the non-monotonic dependence of $m^*$ and $Z$
(more pronounced for smaller $\gamma$), see Fig.~\ref{Effectivem}, i.e.
kuklons are getting lighter as they approach AL.
         
\textit{Conclusions}. 
We performed detailed studies of soliton (kuklon) states in the model with strong quadratic 
electron-phonon interaction by (i) extending previous variational analysis to lattice systems, (ii) revealing two first-order transitions and their properties, which 
explain how kuklons form and subsequently collapse to the single-site states at the variational level, and (iii) solving the problem numerically exactly by the 
X-representation Monte Carlo technique. In exact solutions, transitions are replaced
by smooth crossovers with rapid increase of the effective mass and decrease of the quasiparticle residue  
at strong coupling. An opposite trend is 
discovered when the soliton size is of the order of the lattice spacing. 

Previous numerical work failed to see kuklons because
they form when system parameters take values way beyond 
the limitations of all other unbiased methods. Future work 
should explore properties of bi-polaron states in the same model; bound states are expected to form 
in the same parameter regime where kuklons form and for the same reason---sublinear dependence of the interaction energy on electron density. Figure~\ref{Effectivem} suggests
that it is possible to have light and compact bi-polarons, which is a prerequisite for 
high superconducting temperature at finite density \cite{Chao2,BoseTc1}.

We acknowledge support from the National Science Foundation 
under Grants No.~DMR-2032136 and No.~DMR-2032077.
    
\bibliography{reference}

\begin{thebibliography}{41}%
\makeatletter
\providecommand \@ifxundefined [1]{%
 \@ifx{#1\undefined}
}%
\providecommand \@ifnum [1]{%
 \ifnum #1\expandafter \@firstoftwo
 \else \expandafter \@secondoftwo
 \fi
}%
\providecommand \@ifx [1]{%
 \ifx #1\expandafter \@firstoftwo
 \else \expandafter \@secondoftwo
 \fi
}%
\providecommand \natexlab [1]{#1}%
\providecommand \enquote  [1]{``#1''}%
\providecommand \bibnamefont  [1]{#1}%
\providecommand \bibfnamefont [1]{#1}%
\providecommand \citenamefont [1]{#1}%
\providecommand \href@noop [0]{\@secondoftwo}%
\providecommand \href [0]{\begingroup \@sanitize@url \@href}%
\providecommand \@href[1]{\@@startlink{#1}\@@href}%
\providecommand \@@href[1]{\endgroup#1\@@endlink}%
\providecommand \@sanitize@url [0]{\catcode `\\12\catcode `\$12\catcode
  `\&12\catcode `\#12\catcode `\^12\catcode `\_12\catcode `\%12\relax}%
\providecommand \@@startlink[1]{}%
\providecommand \@@endlink[0]{}%
\providecommand \url  [0]{\begingroup\@sanitize@url \@url }%
\providecommand \@url [1]{\endgroup\@href {#1}{\urlprefix }}%
\providecommand \urlprefix  [0]{URL }%
\providecommand \Eprint [0]{\href }%
\providecommand \doibase [0]{https://doi.org/}%
\providecommand \selectlanguage [0]{\@gobble}%
\providecommand \bibinfo  [0]{\@secondoftwo}%
\providecommand \bibfield  [0]{\@secondoftwo}%
\providecommand \translation [1]{[#1]}%
\providecommand \BibitemOpen [0]{}%
\providecommand \bibitemStop [0]{}%
\providecommand \bibitemNoStop [0]{.\EOS\space}%
\providecommand \EOS [0]{\spacefactor3000\relax}%
\providecommand \BibitemShut  [1]{\csname bibitem#1\endcsname}%
\let\auto@bib@innerbib\@empty
\bibitem [{\citenamefont {Landau}(1933)}]{Landau1}%
  \BibitemOpen
  \bibfield  {author} {\bibinfo {author} {\bibfnamefont {L.}~\bibnamefont
  {Landau}},\ }\href@noop {} {\bibfield  {journal} {\bibinfo  {journal} {Phys.
  Z. Sowjetunion}\ }\textbf {\bibinfo {volume} {3}},\ \bibinfo {pages} {664}
  (\bibinfo {year} {1933})}\BibitemShut {NoStop}%
\bibitem [{\citenamefont {Pekar}(1946)}]{Landau2}%
  \BibitemOpen
  \bibfield  {author} {\bibinfo {author} {\bibfnamefont {S.}~\bibnamefont
  {Pekar}},\ }\href@noop {} {\bibfield  {journal} {\bibinfo  {journal} {Zh.
  Eksp. Teor. Fiz.}\ }\textbf {\bibinfo {volume} {16}},\ \bibinfo {pages} {335}
  (\bibinfo {year} {1946})}\BibitemShut {NoStop}%
\bibitem [{\citenamefont {Landau}\ and\ \citenamefont {Pekar}(1948)}]{Landau3}%
  \BibitemOpen
  \bibfield  {author} {\bibinfo {author} {\bibfnamefont {L.}~\bibnamefont
  {Landau}}\ and\ \bibinfo {author} {\bibfnamefont {S.}~\bibnamefont {Pekar}},\
  }\href@noop {} {\bibfield  {journal} {\bibinfo  {journal} {Zh. Eksp. Teor.
  Fiz.}\ }\textbf {\bibinfo {volume} {18}},\ \bibinfo {pages} {419} (\bibinfo
  {year} {1948})}\BibitemShut {NoStop}%
\bibitem [{\citenamefont {Holstein}(1959)}]{Holstein1}%
  \BibitemOpen
  \bibfield  {author} {\bibinfo {author} {\bibfnamefont {T.}~\bibnamefont
  {Holstein}},\ }\href@noop {} {\bibfield  {journal} {\bibinfo  {journal} {Ann.
  Phys.}\ }\textbf {\bibinfo {volume} {8}},\ \bibinfo {pages} {325} (\bibinfo
  {year} {1959})}\BibitemShut {NoStop}%
\bibitem [{\citenamefont {Holstein}(2000)}]{Holstein2}%
  \BibitemOpen
  \bibfield  {author} {\bibinfo {author} {\bibfnamefont {T.}~\bibnamefont
  {Holstein}},\ }\href@noop {} {\bibfield  {journal} {\bibinfo  {journal} {Ann.
  Phys.}\ }\textbf {\bibinfo {volume} {281}},\ \bibinfo {pages} {725} (\bibinfo
  {year} {2000})}\BibitemShut {NoStop}%
\bibitem [{\citenamefont {Fr\"{o}hlich}\ \emph {et~al.}(1950)\citenamefont
  {Fr\"{o}hlich}, \citenamefont {Pelzer},\ and\ \citenamefont
  {Zienau}}]{Frohlich1}%
  \BibitemOpen
  \bibfield  {author} {\bibinfo {author} {\bibfnamefont {H.}~\bibnamefont
  {Fr\"{o}hlich}}, \bibinfo {author} {\bibfnamefont {H.}~\bibnamefont
  {Pelzer}},\ and\ \bibinfo {author} {\bibfnamefont {S.}~\bibnamefont
  {Zienau}},\ }\href@noop {} {\bibfield  {journal} {\bibinfo  {journal}
  {Philos. Mag.}\ }\textbf {\bibinfo {volume} {41}},\ \bibinfo {pages} {221}
  (\bibinfo {year} {1950})}\BibitemShut {NoStop}%
\bibitem [{\citenamefont {Fr\"{o}hlich}(1954)}]{Frohlich2}%
  \BibitemOpen
  \bibfield  {author} {\bibinfo {author} {\bibfnamefont {H.}~\bibnamefont
  {Fr\"{o}hlich}},\ }\href@noop {} {\bibfield  {journal} {\bibinfo  {journal}
  {Adv. Phys.}\ }\textbf {\bibinfo {volume} {3}},\ \bibinfo {pages} {325}
  (\bibinfo {year} {1954})}\BibitemShut {NoStop}%
\bibitem [{\citenamefont {Franchini}\ \emph {et~al.}(2021)\citenamefont
  {Franchini}, \citenamefont {Reticcioli}, \citenamefont {Setvin},\ and\
  \citenamefont {Diebold}}]{Franchini}%
  \BibitemOpen
  \bibfield  {author} {\bibinfo {author} {\bibfnamefont {C.}~\bibnamefont
  {Franchini}}, \bibinfo {author} {\bibfnamefont {M.}~\bibnamefont
  {Reticcioli}}, \bibinfo {author} {\bibfnamefont {M.}~\bibnamefont {Setvin}},\
  and\ \bibinfo {author} {\bibfnamefont {U.}~\bibnamefont {Diebold}},\
  }\href@noop {} {\bibfield  {journal} {\bibinfo  {journal} {Nature Reviews
  Materials}\ }\textbf {\bibinfo {volume} {6}},\ \bibinfo {pages} {560}
  (\bibinfo {year} {2021})}\BibitemShut {NoStop}%
\bibitem [{\citenamefont {Marchand}\ \emph {et~al.}(2010)\citenamefont
  {Marchand}, \citenamefont {De~Filippis}, \citenamefont {Cataudella},
  \citenamefont {Berciu}, \citenamefont {Nagaosa}, \citenamefont {Prokof'ev},
  \citenamefont {Mishchenko},\ and\ \citenamefont {Stamp}}]{Mona1}%
  \BibitemOpen
  \bibfield  {author} {\bibinfo {author} {\bibfnamefont {D.}~\bibnamefont
  {Marchand}}, \bibinfo {author} {\bibfnamefont {G.}~\bibnamefont
  {De~Filippis}}, \bibinfo {author} {\bibfnamefont {V.}~\bibnamefont
  {Cataudella}}, \bibinfo {author} {\bibfnamefont {M.}~\bibnamefont {Berciu}},
  \bibinfo {author} {\bibfnamefont {N.}~\bibnamefont {Nagaosa}}, \bibinfo
  {author} {\bibfnamefont {N.}~\bibnamefont {Prokof'ev}}, \bibinfo {author}
  {\bibfnamefont {A.}~\bibnamefont {Mishchenko}},\ and\ \bibinfo {author}
  {\bibfnamefont {P.}~\bibnamefont {Stamp}},\ }\href@noop {} {\bibfield
  {journal} {\bibinfo  {journal} {Phys. Rev. Lett.}\ }\textbf {\bibinfo
  {volume} {105}},\ \bibinfo {pages} {266605} (\bibinfo {year}
  {2010})}\BibitemShut {NoStop}%
\bibitem [{\citenamefont {Sous}\ \emph {et~al.}(2018)\citenamefont {Sous},
  \citenamefont {Chakraborty}, \citenamefont {Krems},\ and\ \citenamefont
  {Berciu}}]{Mona2}%
  \BibitemOpen
  \bibfield  {author} {\bibinfo {author} {\bibfnamefont {J.}~\bibnamefont
  {Sous}}, \bibinfo {author} {\bibfnamefont {M.}~\bibnamefont {Chakraborty}},
  \bibinfo {author} {\bibfnamefont {R.}~\bibnamefont {Krems}},\ and\ \bibinfo
  {author} {\bibfnamefont {M.}~\bibnamefont {Berciu}},\ }\href@noop {}
  {\bibfield  {journal} {\bibinfo  {journal} {Phys. Rev. Lett.}\ }\textbf
  {\bibinfo {volume} {121}},\ \bibinfo {pages} {247001} (\bibinfo {year}
  {2018})}\BibitemShut {NoStop}%
\bibitem [{\citenamefont {Xing}\ \emph {et~al.}(2021)\citenamefont {Xing},
  \citenamefont {Chiu}, \citenamefont {Poletti}, \citenamefont {Scalettar},\
  and\ \citenamefont {Batrouni}}]{Scalettar}%
  \BibitemOpen
  \bibfield  {author} {\bibinfo {author} {\bibfnamefont {B.}~\bibnamefont
  {Xing}}, \bibinfo {author} {\bibfnamefont {W.-T.}\ \bibnamefont {Chiu}},
  \bibinfo {author} {\bibfnamefont {D.}~\bibnamefont {Poletti}}, \bibinfo
  {author} {\bibfnamefont {R.}~\bibnamefont {Scalettar}},\ and\ \bibinfo
  {author} {\bibfnamefont {G.}~\bibnamefont {Batrouni}},\ }\href@noop {}
  {\bibfield  {journal} {\bibinfo  {journal} {Phys. Rev. Lett.}\ }\textbf
  {\bibinfo {volume} {126}},\ \bibinfo {pages} {017601} (\bibinfo {year}
  {2021})}\BibitemShut {NoStop}%
\bibitem [{\citenamefont {Zhang}\ \emph {et~al.}(2021)\citenamefont {Zhang},
  \citenamefont {Prokof'ev},\ and\ \citenamefont {Svistunov}}]{Chao1}%
  \BibitemOpen
  \bibfield  {author} {\bibinfo {author} {\bibfnamefont {C.}~\bibnamefont
  {Zhang}}, \bibinfo {author} {\bibfnamefont {N.}~\bibnamefont {Prokof'ev}},\
  and\ \bibinfo {author} {\bibfnamefont {B.}~\bibnamefont {Svistunov}},\
  }\href@noop {} {\bibfield  {journal} {\bibinfo  {journal} {Phys. Rev. B}\
  }\textbf {\bibinfo {volume} {104}},\ \bibinfo {pages} {035143} (\bibinfo
  {year} {2021})}\BibitemShut {NoStop}%
\bibitem [{\citenamefont {Zhang}\ \emph
  {et~al.}(2023{\natexlab{a}})\citenamefont {Zhang}, \citenamefont {Sous},
  \citenamefont {Reichman}, \citenamefont {Berciu}, \citenamefont {Millis},
  \citenamefont {Prokof'ev},\ and\ \citenamefont {Svistunov}}]{Chao2}%
  \BibitemOpen
  \bibfield  {author} {\bibinfo {author} {\bibfnamefont {C.}~\bibnamefont
  {Zhang}}, \bibinfo {author} {\bibfnamefont {J.}~\bibnamefont {Sous}},
  \bibinfo {author} {\bibfnamefont {D.}~\bibnamefont {Reichman}}, \bibinfo
  {author} {\bibfnamefont {M.}~\bibnamefont {Berciu}}, \bibinfo {author}
  {\bibfnamefont {A.}~\bibnamefont {Millis}}, \bibinfo {author} {\bibfnamefont
  {N.}~\bibnamefont {Prokof'ev}},\ and\ \bibinfo {author} {\bibfnamefont
  {B.}~\bibnamefont {Svistunov}},\ }\href@noop {} {\bibfield  {journal}
  {\bibinfo  {journal} {Phys. Rev. X}\ }\textbf {\bibinfo {volume} {13}},\
  \bibinfo {pages} {011010} (\bibinfo {year} {2023}{\natexlab{a}})}\BibitemShut
  {NoStop}%
\bibitem [{\citenamefont {Carbone}\ \emph {et~al.}(2021)\citenamefont
  {Carbone}, \citenamefont {Millis}, \citenamefont {Reichman},\ and\
  \citenamefont {Sous}}]{Sous1}%
  \BibitemOpen
  \bibfield  {author} {\bibinfo {author} {\bibfnamefont {M.}~\bibnamefont
  {Carbone}}, \bibinfo {author} {\bibfnamefont {A.}~\bibnamefont {Millis}},
  \bibinfo {author} {\bibfnamefont {D.}~\bibnamefont {Reichman}},\ and\
  \bibinfo {author} {\bibfnamefont {J.}~\bibnamefont {Sous}},\ }\href@noop {}
  {\bibfield  {journal} {\bibinfo  {journal} {Phys. Rev. B}\ }\textbf {\bibinfo
  {volume} {104}},\ \bibinfo {pages} {L140307} (\bibinfo {year}
  {2021})}\BibitemShut {NoStop}%
\bibitem [{\citenamefont {Sous}\ \emph {et~al.}(2022)\citenamefont {Sous},
  \citenamefont {Zhang}, \citenamefont {Berciu}, \citenamefont {Reichman},
  \citenamefont {Svistunov}, \citenamefont {Prokof'ev},\ and\ \citenamefont
  {Millis}}]{Chao3}%
  \BibitemOpen
  \bibfield  {author} {\bibinfo {author} {\bibfnamefont {J.}~\bibnamefont
  {Sous}}, \bibinfo {author} {\bibfnamefont {C.}~\bibnamefont {Zhang}},
  \bibinfo {author} {\bibfnamefont {M.}~\bibnamefont {Berciu}}, \bibinfo
  {author} {\bibfnamefont {D.}~\bibnamefont {Reichman}}, \bibinfo {author}
  {\bibfnamefont {B.}~\bibnamefont {Svistunov}}, \bibinfo {author}
  {\bibfnamefont {N.}~\bibnamefont {Prokof'ev}},\ and\ \bibinfo {author}
  {\bibfnamefont {A.}~\bibnamefont {Millis}},\ }\href@noop {} {\bibfield
  {journal} {\bibinfo  {journal} {arXiv:2210.14236}\ } (\bibinfo {year}
  {2022})}\BibitemShut {NoStop}%
\bibitem [{\citenamefont {Zhang}\ \emph
  {et~al.}(2023{\natexlab{b}})\citenamefont {Zhang}, \citenamefont
  {Capogrosso-Sansone}, \citenamefont {Boninsegni}, \citenamefont {Prokof'ev},\
  and\ \citenamefont {Svistunov}}]{BoseTc1}%
  \BibitemOpen
  \bibfield  {author} {\bibinfo {author} {\bibfnamefont {C.}~\bibnamefont
  {Zhang}}, \bibinfo {author} {\bibfnamefont {B.}~\bibnamefont
  {Capogrosso-Sansone}}, \bibinfo {author} {\bibfnamefont {M.}~\bibnamefont
  {Boninsegni}}, \bibinfo {author} {\bibfnamefont {N.}~\bibnamefont
  {Prokof'ev}},\ and\ \bibinfo {author} {\bibfnamefont {B.}~\bibnamefont
  {Svistunov}},\ }\href {https://doi.org/10.1103/PhysRevLett.130.236001}
  {\bibfield  {journal} {\bibinfo  {journal} {Phys. Rev. Lett.}\ }\textbf
  {\bibinfo {volume} {130}},\ \bibinfo {pages} {236001} (\bibinfo {year}
  {2023}{\natexlab{b}})}\BibitemShut {NoStop}%
\bibitem [{\citenamefont {Esposito}\ \emph {et~al.}(2017)\citenamefont
  {Esposito}, \citenamefont {Fechner}, \citenamefont {Mankowsky}, \citenamefont
  {Lemke}, \citenamefont {Chollet}, \citenamefont {Glownia}, \citenamefont
  {Nakamura}, \citenamefont {Kawasaki}, \citenamefont {Tokura}, \citenamefont
  {Staub}, \citenamefont {Beaud},\ and\ \citenamefont
  {F\"orst}}]{X2materials1}%
  \BibitemOpen
  \bibfield  {author} {\bibinfo {author} {\bibfnamefont {V.}~\bibnamefont
  {Esposito}}, \bibinfo {author} {\bibfnamefont {M.}~\bibnamefont {Fechner}},
  \bibinfo {author} {\bibfnamefont {R.}~\bibnamefont {Mankowsky}}, \bibinfo
  {author} {\bibfnamefont {H.}~\bibnamefont {Lemke}}, \bibinfo {author}
  {\bibfnamefont {M.}~\bibnamefont {Chollet}}, \bibinfo {author} {\bibfnamefont
  {J.~M.}\ \bibnamefont {Glownia}}, \bibinfo {author} {\bibfnamefont
  {M.}~\bibnamefont {Nakamura}}, \bibinfo {author} {\bibfnamefont
  {M.}~\bibnamefont {Kawasaki}}, \bibinfo {author} {\bibfnamefont
  {Y.}~\bibnamefont {Tokura}}, \bibinfo {author} {\bibfnamefont
  {U.}~\bibnamefont {Staub}}, \bibinfo {author} {\bibfnamefont
  {P.}~\bibnamefont {Beaud}},\ and\ \bibinfo {author} {\bibfnamefont
  {M.}~\bibnamefont {F\"orst}},\ }\href
  {https://doi.org/10.1103/PhysRevLett.118.247601} {\bibfield  {journal}
  {\bibinfo  {journal} {Phys. Rev. Lett.}\ }\textbf {\bibinfo {volume} {118}},\
  \bibinfo {pages} {247601} (\bibinfo {year} {2017})}\BibitemShut {NoStop}%
\bibitem [{\citenamefont {Schilcher}\ \emph {et~al.}(2021)\citenamefont
  {Schilcher}, \citenamefont {Robinson}, \citenamefont {Abramovitch},
  \citenamefont {Tan}, \citenamefont {Rappe}, \citenamefont {Reichman}, ,\ and\
  \citenamefont {Egger}}]{X2materials2}%
  \BibitemOpen
  \bibfield  {author} {\bibinfo {author} {\bibfnamefont {M.}~\bibnamefont
  {Schilcher}}, \bibinfo {author} {\bibfnamefont {P.}~\bibnamefont {Robinson}},
  \bibinfo {author} {\bibfnamefont {D.}~\bibnamefont {Abramovitch}}, \bibinfo
  {author} {\bibfnamefont {L.}~\bibnamefont {Tan}}, \bibinfo {author}
  {\bibfnamefont {A.}~\bibnamefont {Rappe}}, \bibinfo {author} {\bibfnamefont
  {D.}~\bibnamefont {Reichman}}, ,\ and\ \bibinfo {author} {\bibfnamefont
  {D.}~\bibnamefont {Egger}},\ }\href@noop {} {\bibfield  {journal} {\bibinfo
  {journal} {ACS Energy Lett.}\ }\textbf {\bibinfo {volume} {6}},\ \bibinfo
  {pages} {2162} (\bibinfo {year} {2021})}\BibitemShut {NoStop}%
\bibitem [{\citenamefont {Kumar}\ \emph {et~al.}(2021)\citenamefont {Kumar},
  \citenamefont {Yudson},\ and\ \citenamefont {Maslov}}]{X2materials3}%
  \BibitemOpen
  \bibfield  {author} {\bibinfo {author} {\bibfnamefont {A.}~\bibnamefont
  {Kumar}}, \bibinfo {author} {\bibfnamefont {V.~I.}\ \bibnamefont {Yudson}},\
  and\ \bibinfo {author} {\bibfnamefont {D.~L.}\ \bibnamefont {Maslov}},\
  }\href {https://doi.org/10.1103/PhysRevLett.126.076601} {\bibfield  {journal}
  {\bibinfo  {journal} {Phys. Rev. Lett.}\ }\textbf {\bibinfo {volume} {126}},\
  \bibinfo {pages} {076601} (\bibinfo {year} {2021})}\BibitemShut {NoStop}%
\bibitem [{\citenamefont {Nazaryan}\ and\ \citenamefont
  {Feigel'man}(2021)}]{X2materials4}%
  \BibitemOpen
  \bibfield  {author} {\bibinfo {author} {\bibfnamefont {K.~G.}\ \bibnamefont
  {Nazaryan}}\ and\ \bibinfo {author} {\bibfnamefont {M.~V.}\ \bibnamefont
  {Feigel'man}},\ }\href {https://doi.org/10.1103/PhysRevB.104.115201}
  {\bibfield  {journal} {\bibinfo  {journal} {Phys. Rev. B}\ }\textbf {\bibinfo
  {volume} {104}},\ \bibinfo {pages} {115201} (\bibinfo {year}
  {2021})}\BibitemShut {NoStop}%
\bibitem [{\citenamefont {Ngai}(1974)}]{soft1}%
  \BibitemOpen
  \bibfield  {author} {\bibinfo {author} {\bibfnamefont {K.~L.}\ \bibnamefont
  {Ngai}},\ }\href {https://doi.org/10.1103/PhysRevLett.32.215} {\bibfield
  {journal} {\bibinfo  {journal} {Phys. Rev. Lett.}\ }\textbf {\bibinfo
  {volume} {32}},\ \bibinfo {pages} {215} (\bibinfo {year} {1974})}\BibitemShut
  {NoStop}%
\bibitem [{\citenamefont {Kuklov}(1989)}]{Kuklov}%
  \BibitemOpen
  \bibfield  {author} {\bibinfo {author} {\bibfnamefont {A.}~\bibnamefont
  {Kuklov}},\ }\href
  {https://doi.org/https://doi.org/10.1016/0375-9601(89)90154-0} {\bibfield
  {journal} {\bibinfo  {journal} {Physics Letters A}\ }\textbf {\bibinfo
  {volume} {139}},\ \bibinfo {pages} {270} (\bibinfo {year}
  {1989})}\BibitemShut {NoStop}%
\bibitem [{\citenamefont {Kuklov}(1990)}]{Kuklov2}%
  \BibitemOpen
  \bibfield  {author} {\bibinfo {author} {\bibfnamefont {A.~B.}\ \bibnamefont
  {Kuklov}},\ }\href@noop {} {\bibfield  {journal} {\bibinfo  {journal}
  {Superconductivity}\ }\textbf {\bibinfo {volume} {3}},\ \bibinfo {pages}
  {S355} (\bibinfo {year} {1990})},\ \bibinfo {note} {\ ISSN 0235-8964
  [translation from Russian, Sverkhprovodimost (KIAE), 3 (10), 2277
  (1990)]}\BibitemShut {NoStop}%
\bibitem [{\citenamefont {Gogolin}\ and\ \citenamefont
  {Ioselevich}(1991{\natexlab{a}})}]{Gogolin}%
  \BibitemOpen
  \bibfield  {author} {\bibinfo {author} {\bibfnamefont {A.}~\bibnamefont
  {Gogolin}}\ and\ \bibinfo {author} {\bibfnamefont {A.}~\bibnamefont
  {Ioselevich}},\ }\href@noop {} {\bibfield  {journal} {\bibinfo  {journal}
  {JETP letters}\ }\textbf {\bibinfo {volume} {53}},\ \bibinfo {pages} {479}
  (\bibinfo {year} {1991}{\natexlab{a}})}\BibitemShut {NoStop}%
\bibitem [{\citenamefont {Gogolin}\ and\ \citenamefont
  {Ioselevich}(1991{\natexlab{b}})}]{Gogolin2}%
  \BibitemOpen
  \bibfield  {author} {\bibinfo {author} {\bibfnamefont {A.}~\bibnamefont
  {Gogolin}}\ and\ \bibinfo {author} {\bibfnamefont {A.}~\bibnamefont
  {Ioselevich}},\ }\href@noop {} {\bibfield  {journal} {\bibinfo  {journal}
  {JETP letters}\ }\textbf {\bibinfo {volume} {54}},\ \bibinfo {pages} {285}
  (\bibinfo {year} {1991}{\natexlab{b}})}\BibitemShut {NoStop}%
\bibitem [{\citenamefont {Kennes}\ \emph {et~al.}(2017)\citenamefont {Kennes},
  \citenamefont {Wilner}, \citenamefont {Reichman},\ and\ \citenamefont
  {Millis}}]{Andy1}%
  \BibitemOpen
  \bibfield  {author} {\bibinfo {author} {\bibfnamefont {D.}~\bibnamefont
  {Kennes}}, \bibinfo {author} {\bibfnamefont {E.}~\bibnamefont {Wilner}},
  \bibinfo {author} {\bibfnamefont {D.}~\bibnamefont {Reichman}},\ and\
  \bibinfo {author} {\bibfnamefont {A.}~\bibnamefont {Millis}},\ }\href@noop {}
  {\bibfield  {journal} {\bibinfo  {journal} {Nat. Phys.}\ }\textbf {\bibinfo
  {volume} {13}},\ \bibinfo {pages} {479} (\bibinfo {year} {2017})}\BibitemShut
  {NoStop}%
\bibitem [{\citenamefont {Sous}\ \emph {et~al.}(2021)\citenamefont {Sous},
  \citenamefont {Kloss}, \citenamefont {Kennes}, \citenamefont {Reichman},\
  and\ \citenamefont {Millis}}]{Andy2}%
  \BibitemOpen
  \bibfield  {author} {\bibinfo {author} {\bibfnamefont {J.}~\bibnamefont
  {Sous}}, \bibinfo {author} {\bibfnamefont {B.}~\bibnamefont {Kloss}},
  \bibinfo {author} {\bibfnamefont {D.}~\bibnamefont {Kennes}}, \bibinfo
  {author} {\bibfnamefont {D.}~\bibnamefont {Reichman}},\ and\ \bibinfo
  {author} {\bibfnamefont {A.}~\bibnamefont {Millis}},\ }\href@noop {}
  {\bibfield  {journal} {\bibinfo  {journal} {Nat. Comm.}\ }\textbf {\bibinfo
  {volume} {12}},\ \bibinfo {pages} {5803} (\bibinfo {year}
  {2021})}\BibitemShut {NoStop}%
\bibitem [{\citenamefont {van~der Marel}\ \emph {et~al.}(2019)\citenamefont
  {van~der Marel}, \citenamefont {Barantani},\ and\ \citenamefont
  {Rischau}}]{soft2}%
  \BibitemOpen
  \bibfield  {author} {\bibinfo {author} {\bibfnamefont {D.}~\bibnamefont
  {van~der Marel}}, \bibinfo {author} {\bibfnamefont {F.}~\bibnamefont
  {Barantani}},\ and\ \bibinfo {author} {\bibfnamefont {C.~W.}\ \bibnamefont
  {Rischau}},\ }\href {https://doi.org/10.1103/PhysRevResearch.1.013003}
  {\bibfield  {journal} {\bibinfo  {journal} {Phys. Rev. Res.}\ }\textbf
  {\bibinfo {volume} {1}},\ \bibinfo {pages} {013003} (\bibinfo {year}
  {2019})}\BibitemShut {NoStop}%
\bibitem [{\citenamefont {Volkov}\ \emph {et~al.}(2022)\citenamefont {Volkov},
  \citenamefont {Chandra},\ and\ \citenamefont {Coleman}}]{soft3}%
  \BibitemOpen
  \bibfield  {author} {\bibinfo {author} {\bibfnamefont {P.}~\bibnamefont
  {Volkov}}, \bibinfo {author} {\bibfnamefont {P.}~\bibnamefont {Chandra}},\
  and\ \bibinfo {author} {\bibfnamefont {P.}~\bibnamefont {Coleman}},\
  }\href@noop {} {\bibfield  {journal} {\bibinfo  {journal} {Nat. Comm.}\
  }\textbf {\bibinfo {volume} {13}},\ \bibinfo {pages} {4599} (\bibinfo {year}
  {2022})}\BibitemShut {NoStop}%
\bibitem [{\citenamefont {Kiselov}\ and\ \citenamefont
  {Feigel'man}(2021)}]{soft4}%
  \BibitemOpen
  \bibfield  {author} {\bibinfo {author} {\bibfnamefont {D.~E.}\ \bibnamefont
  {Kiselov}}\ and\ \bibinfo {author} {\bibfnamefont {M.~V.}\ \bibnamefont
  {Feigel'man}},\ }\href {https://doi.org/10.1103/PhysRevB.104.L220506}
  {\bibfield  {journal} {\bibinfo  {journal} {Phys. Rev. B}\ }\textbf {\bibinfo
  {volume} {104}},\ \bibinfo {pages} {L220506} (\bibinfo {year}
  {2021})}\BibitemShut {NoStop}%
\bibitem [{\citenamefont {Berciu}(2006)}]{Mona0}%
  \BibitemOpen
  \bibfield  {author} {\bibinfo {author} {\bibfnamefont {M.}~\bibnamefont
  {Berciu}},\ }\href {https://doi.org/10.1103/PhysRevLett.97.036402} {\bibfield
   {journal} {\bibinfo  {journal} {Phys. Rev. Lett.}\ }\textbf {\bibinfo
  {volume} {97}},\ \bibinfo {pages} {036402} (\bibinfo {year}
  {2006})}\BibitemShut {NoStop}%
\bibitem [{\citenamefont {Adolphs}\ and\ \citenamefont
  {Berciu}(2013)}]{Mona00}%
  \BibitemOpen
  \bibfield  {author} {\bibinfo {author} {\bibfnamefont {C.}~\bibnamefont
  {Adolphs}}\ and\ \bibinfo {author} {\bibfnamefont {M.}~\bibnamefont
  {Berciu}},\ }\href@noop {} {\bibfield  {journal} {\bibinfo  {journal}
  {Europhys. Lett.}\ }\textbf {\bibinfo {volume} {102}},\ \bibinfo {pages}
  {47003} (\bibinfo {year} {2013})}\BibitemShut {NoStop}%
\bibitem [{\citenamefont {Adolphs}\ and\ \citenamefont
  {Berciu}(2014)}]{Mona000}%
  \BibitemOpen
  \bibfield  {author} {\bibinfo {author} {\bibfnamefont {C.~P.~J.}\
  \bibnamefont {Adolphs}}\ and\ \bibinfo {author} {\bibfnamefont
  {M.}~\bibnamefont {Berciu}},\ }\href
  {https://doi.org/10.1103/PhysRevB.89.035122} {\bibfield  {journal} {\bibinfo
  {journal} {Phys. Rev. B}\ }\textbf {\bibinfo {volume} {89}},\ \bibinfo
  {pages} {035122} (\bibinfo {year} {2014})}\BibitemShut {NoStop}%
\bibitem [{\citenamefont {Li}\ and\ \citenamefont {Johnston}(2015)}]{detMC1}%
  \BibitemOpen
  \bibfield  {author} {\bibinfo {author} {\bibfnamefont {S.}~\bibnamefont
  {Li}}\ and\ \bibinfo {author} {\bibfnamefont {S.}~\bibnamefont {Johnston}},\
  }\href@noop {} {\bibfield  {journal} {\bibinfo  {journal} {Europhys. Lett.}\
  }\textbf {\bibinfo {volume} {109}},\ \bibinfo {pages} {27007} (\bibinfo
  {year} {2015})}\BibitemShut {NoStop}%
\bibitem [{\citenamefont {Li}\ \emph {et~al.}(2015)\citenamefont {Li},
  \citenamefont {Nowadnick},\ and\ \citenamefont {Johnston}}]{detMC2}%
  \BibitemOpen
  \bibfield  {author} {\bibinfo {author} {\bibfnamefont {S.}~\bibnamefont
  {Li}}, \bibinfo {author} {\bibfnamefont {E.~A.}\ \bibnamefont {Nowadnick}},\
  and\ \bibinfo {author} {\bibfnamefont {S.}~\bibnamefont {Johnston}},\ }\href
  {https://doi.org/10.1103/PhysRevB.92.064301} {\bibfield  {journal} {\bibinfo
  {journal} {Phys. Rev. B}\ }\textbf {\bibinfo {volume} {92}},\ \bibinfo
  {pages} {064301} (\bibinfo {year} {2015})}\BibitemShut {NoStop}%
\bibitem [{\citenamefont {Dee}\ \emph {et~al.}(2020)\citenamefont {Dee},
  \citenamefont {Coulter}, \citenamefont {Kleiner},\ and\ \citenamefont
  {Johnston}}]{detMC3}%
  \BibitemOpen
  \bibfield  {author} {\bibinfo {author} {\bibfnamefont {P.}~\bibnamefont
  {Dee}}, \bibinfo {author} {\bibfnamefont {J.}~\bibnamefont {Coulter}},
  \bibinfo {author} {\bibfnamefont {K.}~\bibnamefont {Kleiner}},\ and\ \bibinfo
  {author} {\bibfnamefont {S.}~\bibnamefont {Johnston}},\ }\href@noop {}
  {\bibfield  {journal} {\bibinfo  {journal} {Commun. Phys.}\ }\textbf
  {\bibinfo {volume} {3}},\ \bibinfo {pages} {145} (\bibinfo {year}
  {2020})}\BibitemShut {NoStop}%
\bibitem [{\citenamefont {Johnston}\ \emph {et~al.}(2013)\citenamefont
  {Johnston}, \citenamefont {Nowadnick}, \citenamefont {Kung}, \citenamefont
  {Moritz}, \citenamefont {Scalettar},\ and\ \citenamefont
  {Devereaux}}]{detMC4}%
  \BibitemOpen
  \bibfield  {author} {\bibinfo {author} {\bibfnamefont {S.}~\bibnamefont
  {Johnston}}, \bibinfo {author} {\bibfnamefont {E.~A.}\ \bibnamefont
  {Nowadnick}}, \bibinfo {author} {\bibfnamefont {Y.~F.}\ \bibnamefont {Kung}},
  \bibinfo {author} {\bibfnamefont {B.}~\bibnamefont {Moritz}}, \bibinfo
  {author} {\bibfnamefont {R.~T.}\ \bibnamefont {Scalettar}},\ and\ \bibinfo
  {author} {\bibfnamefont {T.~P.}\ \bibnamefont {Devereaux}},\ }\href
  {https://doi.org/10.1103/PhysRevB.87.235133} {\bibfield  {journal} {\bibinfo
  {journal} {Phys. Rev. B}\ }\textbf {\bibinfo {volume} {87}},\ \bibinfo
  {pages} {235133} (\bibinfo {year} {2013})}\BibitemShut {NoStop}%
\bibitem [{\citenamefont {Houtput}\ and\ \citenamefont
  {Tempere}(2021)}]{Tempere}%
  \BibitemOpen
  \bibfield  {author} {\bibinfo {author} {\bibfnamefont {M.}~\bibnamefont
  {Houtput}}\ and\ \bibinfo {author} {\bibfnamefont {J.}~\bibnamefont
  {Tempere}},\ }\href {https://doi.org/10.1103/PhysRevB.103.184306} {\bibfield
  {journal} {\bibinfo  {journal} {Phys. Rev. B}\ }\textbf {\bibinfo {volume}
  {103}},\ \bibinfo {pages} {184306} (\bibinfo {year} {2021})}\BibitemShut
  {NoStop}%
\bibitem [{\citenamefont {Feynman}(1955)}]{Feynman}%
  \BibitemOpen
  \bibfield  {author} {\bibinfo {author} {\bibfnamefont {R.~P.}\ \bibnamefont
  {Feynman}},\ }\href {https://doi.org/10.1103/PhysRev.97.660} {\bibfield
  {journal} {\bibinfo  {journal} {Phys. Rev.}\ }\textbf {\bibinfo {volume}
  {97}},\ \bibinfo {pages} {660} (\bibinfo {year} {1955})}\BibitemShut
  {NoStop}%
\bibitem [{\citenamefont {Prokof'ev}\ and\ \citenamefont
  {Svistunov}(2022)}]{Xrep}%
  \BibitemOpen
  \bibfield  {author} {\bibinfo {author} {\bibfnamefont {N.}~\bibnamefont
  {Prokof'ev}}\ and\ \bibinfo {author} {\bibfnamefont {B.}~\bibnamefont
  {Svistunov}},\ }\href@noop {} {\bibfield  {journal} {\bibinfo  {journal}
  {Phys. Rev. B}\ }\textbf {\bibinfo {volume} {106}},\ \bibinfo {pages}
  {L041117} (\bibinfo {year} {2022})}\BibitemShut {NoStop}%
\bibitem [{\citenamefont {Ragni}\ \emph {et~al.}(2023)\citenamefont {Ragni},
  \citenamefont {Hahn}, \citenamefont {Zhang}, \citenamefont {Prokof'ev},
  \citenamefont {Kuklov}, \citenamefont {Klimin}, \citenamefont {Houtput},
  \citenamefont {Svistunov}, \citenamefont {Tempere}, \citenamefont {Nagaosa},
  \citenamefont {Franchini},\ and\ \citenamefont {Mishchenko}}]{X2polaron}%
  \BibitemOpen
  \bibfield  {author} {\bibinfo {author} {\bibfnamefont {S.}~\bibnamefont
  {Ragni}}, \bibinfo {author} {\bibfnamefont {T.}~\bibnamefont {Hahn}},
  \bibinfo {author} {\bibfnamefont {Z.}~\bibnamefont {Zhang}}, \bibinfo
  {author} {\bibfnamefont {N.}~\bibnamefont {Prokof'ev}}, \bibinfo {author}
  {\bibfnamefont {A.}~\bibnamefont {Kuklov}}, \bibinfo {author} {\bibfnamefont
  {S.}~\bibnamefont {Klimin}}, \bibinfo {author} {\bibfnamefont
  {M.}~\bibnamefont {Houtput}}, \bibinfo {author} {\bibfnamefont
  {B.}~\bibnamefont {Svistunov}}, \bibinfo {author} {\bibfnamefont
  {J.}~\bibnamefont {Tempere}}, \bibinfo {author} {\bibfnamefont
  {N.}~\bibnamefont {Nagaosa}}, \bibinfo {author} {\bibfnamefont
  {C.}~\bibnamefont {Franchini}},\ and\ \bibinfo {author} {\bibfnamefont
  {A.}~\bibnamefont {Mishchenko}},\ }\href
  {https://doi.org/10.1103/PhysRevB.107.L121109} {\bibfield  {journal}
  {\bibinfo  {journal} {Phys. Rev. B}\ }\textbf {\bibinfo {volume} {107}},\
  \bibinfo {pages} {L121109} (\bibinfo {year} {2023})}\BibitemShut {NoStop}%
\end{thebibliography}%
\end{document}